\documentstyle[12pt]{article}
\newcommand{\beq}[1]{\begin{equation}\label{#1}}
\newcommand{\eeq}{\end{equation}}
\newcommand{\bear}[1]{\begin{eqnarray}\label{#1}}
\newcommand{\ear}{\end{eqnarray}}
\newcommand{\nn}{\nonumber}

\newcommand{\rf}[1]{(\ref{#1})}

\catcode`\@=11 \@addtoreset{equation}{section}\catcode`\@=12
\newcommand{\be}{\begin{equation}}
\newcommand{\ee}{\end{equation}}
\newcommand{\ba}{\begin{eqnarray}}
\newcommand{\ea}{\end{eqnarray}}
\newcommand{\nl}{ {\hfill \break} }
\newcommand{\np}{ {\newpage } }

\newcommand{\N}{ \mbox{\rm I$\!$N} }
\newcommand{\R}{ \mbox{\rm I$\!$R} }
\def\C{\mbox{\rm {I\kern-.520em C}}}

\newcommand{\eps}{ \varepsilon }

\newcommand{\df}{\mbox{d}}

\newcommand{\absdet}[1]{\vert\det{#1}\vert}
\newcommand{\partlx}[1]{\frac{\partial}{\partial x^{#1}}}
\newcommand{\p}{\partial}

\newcommand{\btu}{\bigtriangleup}
\newcommand{\sq}[1]{\sqrt{|#1|}}

\newcommand\mustbe{\stackrel{!}{=}}
\newcommand\mustbeleq{\stackrel{!}{\leq}}
\begin{document}
%%%%%%%%%%%%%%%%%%%%%%%%%%%%%%%%%%%%%%%%%%%%%%%%%%%%%%%%%%%%%%%%%%%%%%%
%%%%%%%%%%%%%%%%%%%%%%%%%%%%%%%%%%%%%%%%%%%%%%%%%%%%%%%%%%%%%%%%%%%%%%%
%%%%%%%%%%%%%%%%%%%%%%%%%%%%%%%%%%%%%%%%%%%%%%%%%%%%%%%%%%%%%%%%%%%%%%%
%%% Begin of the paper
%%%%%%%%%%%%%%%%%%%%%%%%%%%%%%%%%%%%%%%%%%%%%%%%%%%%%%%%%%%%%%%%%%%%%%%
\centerline{\large \bf 
{Multidimensional $\sigma$-models with composite electric $p$-branes}
\footnote{This work
was financially supported by 
RFBR project grant N 95-02-05785-a, 
DFG grants 436 RUS 113/7, 436 RUS 113/236, and DFG grant Schm 911/6-2.
	   }   
\vspace{1.03truecm}
}

\bigskip

\centerline{\bf \large 
V. D. Ivashchuk, V. N. Melnikov\footnote{e-mail: melnikov@fund.physics.msu.su}
} 
\vspace{0.3truecm}

\centerline{Center for Gravitation and Fundamental Metrology}
\centerline{VNIIMS, 3-1 M. Ulyanovoy Str.}
\centerline{Moscow 117313, Russia}
\vspace{0.95truecm}

\centerline{\bf \large 
M. Rainer\footnote{e-mail: mrainer@aip.de}
}
\vspace{0.3truecm}

\centerline{Gravitationsprojekt, Mathematische Physik I}
\centerline{Institut f\"ur Mathematik, Universit\"at Potsdam}
\centerline{PF 601553, D-14415 Potsdam, Germany}
\vspace{1.03truecm}

\begin{abstract}
\noindent
We consider a gravitational model on a manifold
$M = M_0 \times M_1\times \dots \times M_n$ with oriented  connected 
Einstein internal spaces $M_1, \ldots, M_n$. The matter part of the action 
contains several scalar fields and antisymmetric forms. 
With Ricci-flat internal spaces, the model has a midisuperspace 
representation in form of a $\sigma$-model on $M_0$.
The latter can be used to determine exact composite electric 
$p$-brane solutions, which depend on a set of harmonic functions on $M_0$.
\end{abstract}
\nl 
\section{\bf Introduction}
\setcounter{equation}{0}
%%%%%%%%%%%%%%%%%%%%%%%%%%%%%%%%%%%%%%%%%%%%%%%%%%%%%%%%%%%%%%%%%%%%%%%%
At present there is an increasing interest to investigate 
$p$-brane solutions in the context of multidimensional 
gravity (see e.g. \cite{Dab}-\cite{LP}, and \cite{S} for a review). 
These solutions generalize well-known 
Majumdar-Papapetrou solutions \cite{M,P} 
to the case where the additional matter fields are given by several scalars 
and antisymmetric forms of a generalized electric type.

In this paper we continue investigations \cite{IM4} 
to find new solutions within a nonlinear $\sigma$-model representation 
for the composite case, 
where each antisymmetric electric form is just a sum of 
elementary ones. 
Solutions of such type are by now rather interesting objects in the context 
of $10$- or $11$-dimensional supergravity \cite{CJS,SS}, 
superstring and $M$-theory (cf. \cite{S}, \cite{HT}-\cite{Sch}).

Multidimensional cosmology with Ricci homogeneous factor spaces 
$M_1, \ldots, M_n$ reduces to a dynamical system on minisuperspace 
\cite{IMZ,IM3,Mel,Ra1}. 
In many cases, including multicomponent perfect fluids, 
the effective  system is Toda-like
(see e.g. \cite{GIM,GKMR}). 
This reduction for cosmology
is a special case of multidimensional gravity 
with Ricci homogeneous internal spaces taking 
the effective form  of a nonlinear $\sigma$-model on $M_0$, 
described in \cite{RZ} with a (generalized) synchronous gauge, 
in \cite{IM} with a (generalized) harmonic gauge,
and in \cite{Ra2} with arbitrary gauge. 
Here, we extend the $\sigma$-model representation 
to multidimensional gravity coupled to several scalar matter fields 
and generalized antisymmetric Maxwell fields. 
Under specific orthogonality
conditions for certain minisuperspace vectors 
we obtain exact solutions, below also called
{\em orthobrane} solutions.

It should be noted that, in the following, the metric is apriori 
of arbitrary signature. For simplicity of the dimensional reduction, 
we restrict here from the outset to the case of connected oriented 
Einstein internal spaces.
For the solutions below, global boundary conditions
are hidden implicitly in a set of not further specified harmonic
functions. 
%%%%%%%%%%%%%%%%%%%%%%%%%%%%%%%%%%%%%%%%%%%%%%%%%%%%%%%%%%%%%%%%%%%%%%%%
\section{\bf Multidimensional geometry}
\setcounter{equation}{0}
%%%%%%%%%%%%%%%%%%%%%%%%%%%%%%%%%%%%%%%%%%%%%%%%%%%%%%%%%%%%%%%%%%%%%%%%
Let us now consider a multidimensional manifold
\bear{2.10}
M&=&M_{0}\times\prod_{i=1}^{n} M_{i} ,
\\
D&=&D_0 + \sum_{i=1}^{n} D_i ,
\ear
with $D_i:=\dim M_i \in \N^+$, 
equipped with
a (pseudo) Riemannian metric 
\beq{2.11}
g = e^{2\gamma(x)}g^{(0)}+\sum_{i=1}^{n} e^{2\phi^i(x)}g^{(i)} 
\equiv g_{MN} dz^{M} \otimes dz^{N}
\qquad (\ M,N =1, \ldots, D\ ) 
\eeq
where 
$\phi^i$ and $\gamma$
are smooth scalar functions on $M_0$, 
\beq{2.12}
g^{(0)}  \equiv g^{(0)}_{\mu \nu}(x) dx^{\mu} \otimes dx^{\nu}
\qquad (\ \mu,\nu=1,\ldots, D_{0}\ ) 
\eeq
is an arbitrary (pseudo) Riemannian metric on $M_0$,
for $i=1,\ldots,n$ 
\beq{0a}
g^{(i)} \equiv g_{m_{i} n_{i}}(y_i) dy_i^{m_{i}} \otimes dy_i^{n_{i}}
\qquad (\ m_{i},n_{i}=1,\ldots, D_{i}\ )  
\eeq 
are (pseudo) Riemannian Einstein metrics on $M_i$,
whence there are constants
$\xi_{i}$ such that
\beq{2.13}
R_{m_{i}n_{i}}[g^{(i)}] = \xi_{i} g^{(i)}_{m_{i}n_{i}} .
\eeq
In the following,
we denote by $\btu[g^{(0)}]$ the Laplace-Beltrami operator of $g^{(0)}$,
by $\alpha_{;\mu}$ the covariant derivative of 
$\alpha$ w.r.t. $x^\mu$ and $g^{(0)}$,
by $\alpha_{,\mu}$ its partial derivative w.r.t. $x^\mu$,
and furthermore we set
$(\p\alpha)(\p\beta):=g^{(0)\mu\nu} \alpha^i_{,\mu} \beta^j_{,\nu} $.
Then, the metric (\ref{2.11}) yields
\bear{2.25}
R_{\mu \nu}[g]  
& =  &  
R_{\mu \nu}[g^{(0)} ]  
+ g^{(0)}_{\mu \nu} \Bigl\{
- \btu[g^{(0)}] \gamma +  (2-D_0)  (\p \gamma)^2
- \p \gamma \sum_{j=1}^{n} D_j \p \phi^j 
\Bigr\}
\nn
\\
&  & 
+ (2 - D_0) (\gamma_{;\mu \nu} - \gamma_{,\mu} \gamma_{,\nu})
- \sum_{i=1}^{n} D_i ( \phi^i_{;\mu \nu} - \phi^i_{,\mu} \gamma_{,\nu}
- \phi^i_{,\nu} \gamma_{,\mu} + \phi^i_{,\mu} \phi^i_{,\nu}) ,
\nn\\ 
\label{2.26}
R_{m_{i} n_{i}}[g]  
&= & 
{R_{m_{i}n_{i}}} [g^{(i)}]  
- e^{2 \phi^{i} - 2 \gamma} g^{(i)}_{m_{i}n_{i}}
      \biggl\{ \btu[g^{(0)}] \phi^{i}
+ (\p \phi^{i}) [ (D_0 - 2) \p \gamma  +
	  \sum_{j=1}^{n} D_j \p \phi^j ] \biggr\} ,
\nn
\\
&  & 
i = 1, \ldots, n ,
\ear
and all other Ricci tensor components vanish.
Let us now set   
\beq{2.28}
f \equiv {f}[\gamma, \phi]  := (D_0 - 2) \gamma +\sum_{j=1}^{n} D_j  \phi^j ,
\eeq
where $\phi:=(\phi^i)\in \R^n$
is a vector field composed by the dilatonic scalar fields
of the multidimensional geometry \rf{2.11} on $M$.

Then, the Ricci curvature scalar
decomposes as 
\bear{2.27}
&&R[g] 
=
e^{-2\gamma}R[g^{(0)}]
+\sum_{i=1}^{n} e^{-2\phi^i} R[g^{(i)}] 
\nn\\
& & - e^{-2\gamma}
\left\{
\sum_{i=1}^{n} D_i (\p\phi^i)(\p\phi^i) 
+ (\p f)^2
+(D_0-2) (\p\gamma)^2
+ 2 \btu[g^{(0)}] (f+\gamma)
\right\} , 
\ear
which may be rewritten as
\bear{2.35}
&&R[g] 
=
e^{-2\gamma}R[g^{(0)}]
+\sum_{i=1}^{n} e^{-2\phi^i} R[g^{(i)}] 
\nn\\
& &-  e^{-2\gamma}
\left\{
\sum_{i=1}^{n} D_i (\p\phi^i)(\p\phi^i) 
+ (D_0 - 2) (\p \gamma)^2 - (\p f) \p (f + 2 \gamma) + R_{B} \right\} ,
\ear
\bear{2.36}
&&R_B := (1/\sq {g^{(0)}}) e^{-f}
     \p_{\mu} [2 e^f \sq {g^{(0)} }
     {g^{(0)\mu \nu}} \p_{\nu} (f + \gamma)] ,
\ear
where the last term will be seen below to yield just a boundary
contribution. 
Here we set $|g| := |\det (g_{MN})|$, 
$|g^{(0)}| := |\det (g^{(0)}_{\mu\nu})|$, and analogously for all metrics 
$g^{(i)}$, $i=1, \ldots, n$.

For $i=1,\ldots,n$, we assume all $M_{i}$ to be 
connected and oriented.
A volume form on $M_i$ is given by
\beq{2.14}
\tau_i  := \sqrt{|g^{(i)}(y_i)|}
\ dy_i^{1} \wedge \ldots \wedge dy_i^{D_i} ,
\eeq
and the total internal space volume is
\beq{2.32}
V:= \prod_{i=1}^n V_i, \quad
V_i := \int_{M_i} \tau_i =\int_{M_i} d^{D_i}y_i \sq{g^{(i)} } .
\eeq
In any case, we tune the coupling constant $\kappa$ such that
\beq{vol}
\kappa_0:=\kappa \cdot V^{-\frac{1}{2}}  
\eeq
becomes the physical coupling constant 
(for the case $V\to\infty$, $\kappa\to\infty$,
see also \cite{RZ})
for the effective model on $M_0$ in Einstein frame (cf. \rf{3.2} below). 

With \rf{2.36}, we define a purely geometrical action on $M$ by   
\beq{2.37}
S_{\rm GHY} := \frac{1}{2\kappa^{2}} \int_{M} d^{D}z \sq g
     \{ e^{-2 \gamma} R_{B} \} .
\eeq
We may use \rf{2.32} and \rf{vol} 
to see that \rf{2.37} is just a pure boundary term,
given by a total flow through $\p M$.
In fact
\beq{S_GH}
S_{GHY} = 
\frac{1}{\kappa^2_0}
\int_{M_0}\df^{D_0}x 
\partlx{\lambda}
\left( 
e^{f} 
\sqrt{\absdet{g^{(0)}}} g^{(0)\lambda\nu}
\partlx{\nu} (f+\gamma) 
\right) 
\eeq
is indeed a (generalized) Gibbons-Hawking-York  type action
(compare \cite{GH,Y}),
which here is given by an effective $D_0$-dimensional flow 
through $\p M_0$.

Furthermore, let
\bear{2.16}
\Omega \equiv \Omega(n):={\cal P}^{\{ 1,2  \ldots, n \}}_*
\equiv \{ \{ 1 \}, \{ 2 \}, \ldots, \{ n \},
\{ 1,2 \}, \ldots, \{ 1,2  \ldots, n \} \} 
\ear
be the set of all non-empty
subsets of $\{ 1, \ldots,n \}$.
The number of elements in $\Omega$ is $|\Omega| = 2^n - 1$.
Then, for any $I = \{ i_1, \ldots, i_k \} \in \Omega$, we may assume
without restriction $i_1 < \ldots < i_k$,
and define 
\bear{2.20}
M_{I} &:=& \prod_{i\in I} M_{i}= M_{i_1}  \times  \ldots \times M_{i_k},
\\
\label{2.19}
D_I &:=&  \sum_{i \in I} D_i = D_{i_1} + \ldots + D_{i_k},
\\
\label{2.16a}
\tau_I &:=& \tau_{i_1}  \wedge \ldots \wedge \tau_{i_k} .
\ear
We also consider
the dimension as a function of $I\subset\{ 1, \ldots,n \}$, defined by 
\bear{dimfunc}
D(I):=&D_I, \qquad& I\in\Omega ,
\\
D(I):=&0 , \qquad& I=\emptyset .
\ear
Furthermore  on $M_I$ we 
define a signature parameter 
\beq{2.31}
\varepsilon(I) := \prod_{i\in I} 
{\rm sign}( {\rm det } (g^{(i)}_{m_{i}n_{i}})) 
\in \{\pm 1\} ,
\eeq
and the topological numbers
\beq{3.8}
l_{jI} := - \sum_{i \in I} D_i \delta^i_j ,
\qquad j =1,\ldots,n .
\eeq
%%%%%%%%%%%%%%%%%%%%%%%%%%%%%%%%%%%%%%%%%%%%%%%%%%%%%%%%%%%%%%%%%%%%%%%%
\section{\bf Matter fields}
\setcounter{equation}{0}
%%%%%%%%%%%%%%%%%%%%%%%%%%%%%%%%%%%%%%%%%%%%%%%%%%%%%%%%%%%%%%%%%%%%%%%%
Here we consider matter fields of scalar and generalized electric type.
Here all fields will be essentially homogeneous in the internal space
$M_{1} \times \ldots \times M_{n}$.
 
The part of scalar matter, which is homogeneous on $M$, is described by a
cosmological constant $\Lambda$.
Besides, on $M$ there may be several scalar matter fields  
\beq{2.24}
\varphi^\alpha = \varphi^\alpha(x), \qquad x\in M_0, 
\qquad \alpha=1,\ldots,l \in \N ,
\eeq
which are homogeneous in the internal space 
$M_{1} \times \ldots \times M_{n}$ and
hence are effectively just just depend on $M_0$.
The target values of these fields are collected in a vector 
$\varphi(x)=(\varphi^\alpha(x))\in\R^l$. 
We assume further, that there exists
a non-degenerate, symmetric 
$l\times l$ matrix
$(C_{\alpha\beta})$ which yields a metric on the target space $\R^l$ of 
the scalar matter fields.  $(C^{\alpha \beta})$ denotes the inverse of 
$(C_{\alpha \beta})$.

Given  some finite set $\Delta$, we define, for any
$a \in \Delta$,
a linear $1$-form $\lambda_{a}$ on $\R^l$ with 
$\lambda_{a} (\varphi)=\lambda_{a \alpha}\varphi^\alpha$, and we 
set $\lambda^{\alpha}_{a} := \lambda_{a\beta} C^{\beta\alpha}$. 
Moreover, for any $a \in \Delta$,
we define
a differential $n_a$-form ($n_a \geq 2$) on $M$ with
\beq{2.2}
F^a =
\frac{1}{n_a!} F^a_{M_1 \ldots M_{n_a}}
dz^{M_1} \wedge \ldots \wedge dz^{M_{n_a}} 
\qquad (\ M_{1},\ldots, M_{n_a}=1,\ldots, D\ ) ,  
\eeq
and we denote
\beq{2.3}
(F^a)^2 :=
g^{M_1 N_1} \ldots g^{M_{n_a} N_{n_a}}
F^a_{M_1 \ldots M_{n_a}} F^a_{N_1 \ldots N_{n_a}} .
\eeq
Below we specialize the components $F^a$ to a generalized electric type.
Let $\Delta_e \subset \Delta$ be a non-empty subset,
and 
\bear{2.e}
j_e: \Delta_e &\to& {\cal P}^{\Omega}_* 
\nn \\
\qquad  a &\mapsto& \Omega_{a,e} \neq \emptyset 
\ear
a map from $\Delta_e$ into the set ${\cal P}^{\Omega}_*$
of all non-empty subsets of $\Omega$ 
such that
\beq{2.23}
D(I) = n_a -1 , 
\qquad a \in \Delta_e, \qquad I \in \Omega_{a,e} .
\eeq
With the map \rf{2.e} we make the following ansatz for 
generalized electric potential forms $A^a$ with $F^a=dA^a$, 
$a \in \Delta_e $: 
For any $a\in\Delta_e$ and $I \in \Omega_{a,e}$ 
there shall exist a smooth potential function $\Phi^{a,I}$, 
which depends just on $M_0$, 
thus being homogeneous on the whole internal space $M_I$. 
The target values of these potential fields are collected in a multi-vector 
$\Phi(x)=(\Phi^{a,I}(x))\in 
\oplus_{{a\in \Delta_e}} \R^{| \Omega_{a,e}|}$. 
We then set
\bear{2.e1}
A^a:=&\sum_{I \in \Omega_{a,e}} A^{a,I} ,  \qquad & a \in \Delta_e ,
\\
\label{2.e2}
A^a:=&0 ,  \qquad &  a \in \Delta \setminus \Delta_e ,
\ear
where, with $\tau_{I}$ from (\ref{2.16a}),
\beq{2.17}
A^{a,I} := \Phi^{a,I} (x) \tau_{I} ,  
\qquad a \in \Delta_e, \qquad I \in \Omega_{a,e} ,
\eeq
(no summation over $I$) are elementary, electric type potential forms.
In components, relation (\ref{2.17}) reads
\bear{2.18}
A^{a,I}_{P_1 \ldots P_{D(I)}}(x,y) =
\varepsilon_{P_1 \ldots P_{D(I)}}
\sqrt{|g^{(i_1)}(y_{i_1})|} \ldots \sqrt{|g^{(i_k)}(y_{i_k})|} \ 
\Phi^{a,I}(x) , 
\ear 
where indices $P_1, \ldots, P_{D(I)}$   correspond to $M_I$
(see \rf{2.20}) with $I \in \Omega_{a,e}$, $a \in \Delta_e$.

Obviously the choice \rf{2.e1}-\rf{2.17} satisfies \rf{2.23} and thus 
corresponds consistently to a map \rf{2.e}, whence
the generalized electric $n_a$-forms $F^a$ read
\bear{2.e3}
F^a=&\sum_{I \in \Omega_{a,e}} F^{a,I} ,  \qquad & a \in \Delta_e ,
\\
\label{2.e4}
F^a=&dA^a=0 ,   \qquad & a \in \Delta \setminus \Delta_e , 
\ear
where
\beq{2.21}
F^{a,I} = dA^{a,I} = d \Phi^{a,I} \wedge \tau_{I} ,
\eeq
or, in components,
\bear{2.22}
F^{a,I}_{\mu P_1 \ldots P_{D(I)}} =
- F^{a,I}_{P_1 \mu \ldots P_{D(I)}} = \ldots =
%\\ \nn
\varepsilon_{P_1 \ldots P_{D(I)}} 
\sqrt{|g^{(i_1)}|} \ldots \sqrt{|g^{(i_k)}|} \ \p_{\mu} \Phi^{a,I}  ,
\ear
for $a \in \Delta_e$, $I\in \Omega_{a,e}$.

Note that \rf{2.e2} is satisfied trivially by
constant gauge potentials  $\Phi^{{a}} \in\R$,
$a \in  \Delta \setminus  \Delta_{e}$.
Neglecting trivial field components \rf{2.e4}, 
we can always set $\Delta=\Delta_e$.
%%%%%%%%%%%%%%%%%%%%%%%%%%%%%%%%%%%%%%%%%%%%%%%%%%%%%%%%%%%%%%%%%%%%%%%%
\section{\bf The multidimensional $\sigma$-model}
\setcounter{equation}{0}
%%%%%%%%%%%%%%%%%%%%%%%%%%%%%%%%%%%%%%%%%%%%%%%%%%%%%%%%%%%%%%%%%%%%%%%%
For the multidimensional geometry and the matter fields defined
in the previous sections, 
we consider the action
\bear{2.1}
S = \frac{1}{2\kappa^{2}}
\int_{M} d^{D}z \sqrt{|g|} \{ {R}[g] - 2 \Lambda - C_{\alpha\beta}
g^{MN} \partial_{M} \varphi^\alpha \partial_{N} \varphi^\beta
\nn\\ 
- \sum_{a \in \Delta}
\frac{1}{n_a!} \exp[ 2 \lambda_{a} (\varphi) ] (F^a)^2 \}
+ S_{GHY} .
\ear
This is an action of a self-gravitating $\sigma$ model on  $M$, with 
topological term $S_{GHY}$ and $l$-dimensional (scalar field) target space,
coupled to generalized electric fields on $D_I$-dimensional hypersurfaces
$M_I$ in $M$.

Variation of (\ref{2.1}) results in the 
equations of motion
\bear{2.4}
R_{MN} - \frac{1}{2} g_{MN} R  &=&   T_{MN} - \Lambda g_{MN} ,
\\
\label{2.5}
{\btu}[g] \varphi^\alpha -
\sum_{a \in \Delta} \frac{\lambda^{\alpha}_a}{n_a!}
e^{2 \lambda_{a}(\varphi)} (F^a)^2 &=& 0 ,
\\
\label{2.6}
\nabla_{M_1}[g] (e^{2 \lambda_{a}(\varphi)}
F^{a, M_1 \ldots M_{n_a}})  &=&  0 ,
\ear
$a \in \Delta$, $\alpha=1,\ldots,l$.
In (\ref{2.4}) the $D$-dimensional energy-momentum 
resulting from \rf{2.1} is given by a sum
\bear{2.7}
T_{MN} :=  \sum_{\alpha=1}^l T_{MN}[\varphi^\alpha,g]
+ \sum_{a\in\Delta} e^{2 \lambda_{a}(\varphi)} T_{MN}[F^a,g] ,
\ear
of contributions from scalar and
generalized electric matter fields,
\bear{2.8}
T_{MN}[\varphi^\alpha,g] &:=&
\p_{M} \varphi^\alpha \p_{N} \varphi^\alpha -
\frac{1}{2} g_{MN} \p_{P} \varphi^\alpha \p^{P} \varphi^\alpha ,
\\
\label{2.9}
T_{MN}[F^a,g] &:=&
\frac{1}{n_{a}!}  \left[ 
- \frac{1}{2} g_{MN} (F^{a})^{2}
 + n_{a}  F^{a}_{\ M M_2 \ldots M_{n_a}} F_{\ N}^{a\ M_2 \ldots M_{n_a}}
\right] .
\ear
In the following, we will only consider the case where 
the tensor  \rf{2.7}
is without further constraints block-diagonal, in agreement with
the multidimensional decomposition of the geometry  (\ref{2.26})
and the field equations \rf{2.4}. 
This condition can be violated  only by the second term
of  \rf{2.9}, namely when in 
$F^{a}_{\ M M_2 \ldots M_{n_a}} F_{\ N}^{a\ M_2 \ldots M_{n_a}}$
the indices $M$ and $N$ take values in different $1$-dimensional
factor spaces. 
We give now some criteria for block-diagonality of \rf{2.9}.
\nl
\noindent
{\bf Proposition 1:}
A sufficient technical criterion for block-diagonality
of \rf{2.7}, without further restrictions on the $p$-branes, is
\beq{2.r1}
n_1:=| W_1 | \mustbeleq 1 , \quad W_1:= \{ i \mid i>0,\ D_i=1\} .
\eeq
Condition \rf{2.r1} just demands the number of (internal) $1$-dimensional 
manifolds $M_i$, $i>0$, in \rf{2.10} not to be more than $1$.
A less restrictive criterion can be given, if we put some 
additional condition on the $p$-brane  intersections.
\nl
\noindent
{\bf Proposition 2:}
Block-diagonality is still satisfied
without further constraints on \rf{2.7}, 
if, for $n_1>1$, the $p$-branes satisfy 
for all $a\in \Delta_e$, $i,j\in W_1$ with $i\neq j$, 
the condition 
\beq{2.r2}
W(a;i,j):= \{ (I,J) \mid I,J\in \Omega_{a,e},\ 
			(I\cap J) \cup \{i\} = I \not\ni j,\ 
			(I\cap J) \cup \{j\} = J \not\ni i  \}  
\mustbe \emptyset .
\eeq
Cases with non-trivial constraints on the tensor \rf{2.7}
are considered e.g. in \cite{IKR}.
Here we restrict to the case where, 
via condition \rf{2.r1} or  \rf{2.r2},
the tensor  \rf{2.7} is automatically block-diagonal.
Then, the decomposition of the multidimensional
curvature according to (\ref{2.26}) and (\ref{2.35}) may be used
to verify that, for 
a multidimensional geometry (\ref{2.11}) with internal Einstein spaces
\rf{2.13}, with scalar matter (\ref{2.24}) and generalized electric
matter of (\ref{2.e3})-(\ref{2.21}),
the field equations
(\ref{2.4})-(\ref{2.6})  
can be obtained as just
equations of motion on $M_0$ from an action
\bear{2.29}
S_{f} \equiv S_{f}[g^{(0)} , \gamma,\phi,\varphi,\Phi]
 :=   \frac{1}{2 \kappa^{2}_0}
     \int_{M_0} d^{D_0}x   \sq {g^{(0)}}
     e^{f(\gamma, \phi)} \biggl\{ {R}[g^{(0)} ]
- \sum_{i =1}^{n} D_i (\p \phi^i)^2
\nn\\ 
 -  (D_0 - 2) (\p \gamma)^2   + (\p f) \p (f + 2\gamma)
+      \sum_{i=1}^{n} \xi_{i}D_i e^{2 (\gamma- \phi^i ) } -
     2 \Lambda e^{2 \gamma} - {\cal L}  \biggr\},
\ear
where
\bear{2.30}
{\cal L} \equiv {\cal L}[g^{(0)},\phi,\varphi,\Phi]
:= C_{\alpha\beta} \p \varphi^\alpha \p \varphi^\beta
+ \sum_{a \in \Delta_e}\sum_{I \in \Omega_{a,e}}
\varepsilon(I) e^{2 (\lambda_a(\varphi) - \sum_{i \in I} D_i \phi^i )}
(\p \Phi^{a,I})^2 .
\ear
Indeed
the action (\ref{2.29}) coincides with the action (\ref{2.1}), i.e.
\beq{2.33}
  S_{f}[g^{(0)} ,\gamma,\phi,\varphi,\Phi] =
  S[g(g^{(0)},\gamma,\phi), \varphi, F(\Phi)] ,
\eeq
where $g \equiv g(g^{(0)},\gamma,\phi)$  and $F \equiv F(\Phi)$
are defined by the relations (\ref{2.11}) and (\ref{2.21})
respectively. 
Note that the second term in (\ref{2.30}) appears,
according to (\ref{2.22}), due to the relation 
\beq{2.38}
\frac{1}{n_a!} (F^{a,I})^2
= \varepsilon(I)
e^{- 2 (\gamma  + \sum_{i \in I} D_i \phi^i)} (\p \Phi^{a,I})^2 
\eeq
with $a \in \Delta_e$, $I \in \Omega_{a,e}$.

In \rf{2.29}, a non-trivial scalar field  
$f\equiv {f}[\gamma(\phi), \phi]\neq 0$ 
yields a non-minimal
coupling of the internal spaces (logarithmic) dilation vector $\phi$ 
to the geometry on $M_0$,
whence this geometry does not define an Einstein frame.
In such a case one might conformally transform the model
to obtain an Einstein frame. This was e.g. discussed
for the matter free case in \cite{RZ}. 
Alternatively, we might try to choose a gauge
such that $f= 0$.  
Then, 
all fields are minimally coupled to the geometry on $M_0$, 
and (\ref{2.29}) becomes an Einsteinian non-linear $\sigma$-model on $M_0$. 

Here we will not discuss further
the subtle relation between different gauges of $f$ and different 
conformally related frames (see e.g. \cite{Ra1,Ra2}).
Rather we assume in the following $D_0\neq 2$, and use (like in \cite{IM})
the (generalized) harmonic gauge  $f=0$, which then exists with
\beq{3.1}
\gamma \equiv {\gamma}(\phi) =
\frac{1}{2- D_0}  \sum_{i =1}^{n} D_i \phi^i .
\eeq
For the usual cosmological case  with $D_0 =1$ and  
$g^{(0)} = - dt \otimes dt$,
the gauge (\ref{3.1}) is of course the harmonic time gauge
(cf. \cite{IMZ,Ra1}).

With $\gamma\equiv\gamma(\phi)$ given by \rf{3.1}, 
the multi-scalar-tensor action (\ref{2.29})
simplifies to that of a purely Einsteinian $\sigma$-model 
\bear{3.2}
S_{0}[g^{(0)} ,\phi, \varphi,\Phi] &\equiv&
S_{0}[g^{(0)},\gamma(\phi),\phi,\varphi,\Phi] 
\nn\\
&=&
\frac{1}{2 \kappa^{2}_0}
     \int_{M_0} d^{D_0}x \sq {g^{(0)} }
\Bigl\{
{R}[g^{(0)}]
\\ \nn
&&- G_{ij} g^{{(0)}\ \mu \nu} \p_{\mu} \phi^i  \p_{\nu} \phi^j -
2 {V}(\phi) - {\cal L}
\Bigr\} .
\ear
Here
\beq{3.3}
G_{ij} := D_i \delta_{ij} + \frac{D_i D_j}{D_0 -2}, 
\qquad i, j = 1, \ldots, n ,
\eeq
yields a midisuperspace metric on
the purely dilatonic part of the $\sigma$-model
(i.e. a minisuperspace metric on the target space ${\R}^{n}$ of $\phi$), 
and
\beq{3.4}
 V \equiv {V}(\phi)
:= e^{2 {\gamma}(\phi)} 
\{ \Lambda -\frac{1}{2}   \sum_{i =1}^{n} \xi_i D_i e^{-2 \phi^i} \}
\eeq
is a potential on the dilatonic target minisuperspace ${\R}^{n}$.

Thus, \rf{3.2} is the action of a self-gravitating
$\sigma$ model on  $M_0$ with a
$(n + l + \sum_{a\in \Delta_e}| \Omega_{a,e}|)$-dimensional target space
and a self-interaction of the dilatonic vector field $\phi$ described by the
potential (\ref{3.4}).
%%%%%%%%%%%%%%%%%%%%%%%%%%%%%%%%%%%%%%%%%%%%%%%%%%%%%%%%%%%%%%%%%%%%%%%%
\section{Effective $\sigma$-model with zero dilatonic potential} 
\setcounter{equation}{0}
%%%%%%%%%%%%%%%%%%%%%%%%%%%%%%%%%%%%%%%%%%%%%%%%%%%%%%%%%%%%%%%%%%%%%%%%
Let us now consider the special case $\xi_1 =\ldots=\xi_2 = \Lambda = 0$,
where for $i = 1, \ldots, n$ the spaces $(M_i, g^{(i)})$ are Ricci-flat,
and the cosmological constant is zero.
In this case the potential  (\ref{3.4}) is trivial $V =0$,
and the self-interaction of the internal spaces dilatonic vector $\phi$
vanishes. 

With $N := n + l$, we define
an $N$-dimensional vector field
$(\sigma^A) := (\phi^i, \varphi^{\alpha})$, $A=1,\ldots,n,n+1,\ldots,N$,
composed by dilatonic and matter scalar fields, and
a non-degenerate (block-diagonal) $N \times N$-matrix
\beq{3.6}
\hat{G} = \left(\hat{G}_{AB} \right)    =
			     \left(
			      \begin{array}{cc}
 
			       G_{ij} &  0 \\
				   0   &  C_{\alpha \beta}
			       \end{array}
			  \right) .
\eeq
Furthermore we define a multi-index set as 
\beq{3.s}
S_e  := \bigsqcup_{a \in \Delta_e} \{a\} \times\Omega_{a,e} ,
\eeq
and, 
for $s=(a,I)\in S_e$, i.e. $a\in\Delta_e$ and $I \in \Omega_{a,e}$,
we set
$\eps_s:= \eps(I)$, $\Phi^s:=\Phi^{a,I}$,
and define a $N \times |S_e|$-matrix 
\beq{3.7}
L = \left(L_{As} \right)
=
\left( \begin{array}{cc}
	    &L_{i s} \\
	     &L_{\alpha s}
	     \end{array}
\right)
:=
\left( \begin{array}{cc}
	    &l_{i I} \\
	     &\lambda_{\alpha a}
	     \end{array}
\right) ,
\eeq
where $l_{i I}$ are the topological numbers from \rf{3.8}, 
$i =1,\ldots,n$, and 
$\lambda_{\alpha a}$ are the $\phi^{\alpha}$-components of the linear forms
$\lambda_a(\phi)$, $\alpha=1,\ldots,l$. 
With these definitions, \rf{3.2} takes the form
\bear{3.5}
S_{0} = \frac{1}{2 \kappa^{2}_0} 
\int_{M_0} d^{D_0}x \sq {g^{(0)} } \Bigl\{ {R}[g^{(0)}]
- \hat{G}_{AB} \p \sigma^A \p \sigma^B
- \sum_{s \in S_e} \varepsilon_s e^{2 L_{A s} \sigma^A}
(\p \Phi^s)^2  \Bigr\} 
\nn\\
\qquad (\ A,B=1,\ldots, N\ ) .
\ear
The corresponding equations of motion from \rf{3.5} are
\bear{3.9}
R_{\mu \nu}[g^{(0)}]  =
\hat{G}_{AB} \p_{\mu} \sigma^A \p_{\nu} \sigma^B
+ \sum_{s \in S_e} \eps_s e^{2 L_{As} \sigma^A}
\p_{\mu} \Phi^s  \p_{\nu} \Phi^s ,
&\qquad&
\mu,\nu=1,\ldots, D_0 ,
\\
\label{3.10}
\hat{G}_{AB} {\btu}[g^{(0)}] \sigma^B
-  \sum_{s \in S_e} \eps_s L_{As} e^{2 L_{Cs} \sigma^C} (\p \Phi^s)^2 = 0 , 
&\qquad&
A = 1,\ldots, N ,
\\ \label{3.11}
\p_{\mu} \left( \sqrt{|g^{(0)}|} g^{{(0)} \mu \nu} e^{2 L_{As}
\sigma^A} \p_{\nu} \Phi^s \right) = 0 , 
&\qquad&
s \in S_e .
\ear 
%%%%%%%%%%%%%%%%%%%%%%%%%%%%%%%%%%%%%%%%%%%%%%%%%%%%%%%%%%%%%%%%%%%%%%%%
\section{Exact solutions}
\setcounter{equation}{0}
%%%%%%%%%%%%%%%%%%%%%%%%%%%%%%%%%%%%%%%%%%%%%%%%%%%%%%%%%%%%%%%%%%%%%%%%
Now we present a special class of solutions of the field equations
(\ref{3.9})-(\ref{3.11}). Since $\hat{G}$ is regular, $\hat{G}^{-1}$ exists.
We set $(\hat{G}^{AB}) := (\hat{G}_{AB})^{-1}$. 
\beq{3.21}
< X,Y > := X_A \hat{G}^{AB} X_{B} .
\eeq
For $s\in S_e$ let us now consider vectors 
\beq{3.20}
L_{s} = (L_{As}) \in \R^N . 
\eeq
\nl
\noindent
{\bf Definition:}
A non-empty set $S_{*} \subset S_e$ is called an {\em orthobrane} index set,
iff there exists a family of real non-zero coefficients 
$\{\nu_s\}_{s \in S_*}$, such that 
\beq{3.19}
< L_s,L_r > = 
(L^{T} \hat{G}^{-1} L)_{sr} = - \eps_s (\nu_s)^{-2}\delta_{sr} ,
\qquad s,r \in S_* .
\eeq
For  $s \in S_{*}$  and  $A = 1, \ldots, N$, we set
\beq{3.17}
\alpha^A_s := - \eps_s (\nu_s)^{2}  \hat{G}^{AB} L_{Bs} .
\eeq
{}\nl
Here, (\ref{3.19}) is just an orthogonality condition for the
vectors $L_s$, $s \in S_{*}$.
Note that $< L_{s}, L_{s} >$ has just the opposite sign of
$\eps_s$, $s \in S_{*}$.
With the definition above, 
we obtain a criterion for the existence of solutions. 
\nl
\noindent
{\bf Theorem:}
Let  $ S_{*}$ be an orthobrane index set
with coefficients \rf{3.17}.
If for any  $s\in S_{*}$
there is a function  $H_s > 0$ on $M_0$ such that 
\beq{3.18}
{\btu}[g^{(0)}] H_s = 0 ,
\eeq
i.e. $H_s$ is harmonic on $M_0$, 
then, 
the field configuration
\bear{3.13}
R_{\mu \nu}[g^{(0)}]  = 0 ,  
&\qquad&
\mu,\nu=1,\ldots, D_0 ,
\\
\label{3.14}
\sigma^A = \sum_{s \in S_*} \alpha^A_s \ln H_s ,
&\qquad&
A = 1, \ldots, N ,
\\
\label{3.15}
\Phi^s  = \frac{\nu_s}{H_s} ,
&\qquad&
s \in S_{*} ,
\\
\label{3.16}
\Phi^{{s}} = C_{s} ,
&\qquad&
{s} \in  S_e \setminus  S_{*} ,
\ear
with constants $C_{s}\in\R$,
satisfies the field equations (\ref{3.9})-(\ref{3.11}).
\hfill\mbox{$\Box$}\break
\nl
This theorem follows just from substitution of 
(\ref{3.19})-(\ref{3.16}) into
the equations of motion (\ref{3.9})-(\ref{3.11}).
For the following, we assume that the set $\Delta_e$ and 
the map $j_e$ in \rf{2.e} are chosen such that $S_e$
is an orthobrane index set, and we set $S_{*}=S_{e}$, 
whence \rf{3.16} becomes empty.
{}From (\ref{3.6}), (\ref{3.7}) and
(\ref{3.21}) we get
\beq{3.22}
< L_{s}, L_{r} > =  G^{ij} l_{iI} l_{jJ}
+ C^{\alpha\beta} {\lambda}_{\alpha a}  {\lambda}_{\beta b} ,
\eeq
with $s=(a,I)$ and $r=(b,J)$ in $S_e$
($a,b\in\Delta_e$, $I \in \Omega_{a,e}$,  $J \in \Omega_{b,e}$).
Here, the inverse of the dilatonic midisuperspace metric (\ref{3.3})
is given by
\beq{3.26}
G^{ij} =
\frac{\delta_{ij}}{D_i} + \frac{1}{2 - D} ,
\eeq
whence, for $I, J \in \Omega$, with topological numbers $l_{iI}$
{}from \rf{3.8}, we obtain   
\beq{3.27}
 G^{ij} l_{iI} l_{jJ} = D(I \cap J) + \frac{D(I) D(J)}{2-D} ,
\eeq
which is again a purely topological number.

We set $\nu_{a,I} := \nu_{(a,I)}$. Then,
due to (\ref{3.22}) and (\ref{3.27}),
the orthobrane condition (\ref{3.19}) reads
\beq{3.28}
D(I \cap J) + \frac{D(I) D(J)}{2-D}
+ C^{\alpha\beta} {\lambda}_{\alpha a}  {\lambda}_{\beta b}
= - \eps(I) (\nu_{a,I})^{-2} \delta_{ab}\delta_{I,J} ,
\eeq
for $a, b \in \Delta_{e}$, $I \in \Omega_{a,e}$, $I \in \Omega_{b,e}$. 
With $(a,I)=s\in S_e$, the coefficients (\ref{3.17}) are  
%with $\alpha^A_{a,I} := \alpha^A_{(a,I)}$ 
\bear{3.29}
\alpha^i_s
&=& - \eps(I) G^{ij}l_{jI}  \nu_{a,I}^{2}
= \eps(I) 
\biggl( \sum_{j \in I} \delta^i_j  + \frac{D(I)}{2-D} \biggr)
\nu_{a,I}^{2} ,
\qquad i = 1, \ldots, n ,
\\
\label{3.30}
\alpha^\beta_s &=&
- \eps(I) C^{\beta\gamma}\lambda_{\gamma a}  \nu_{a,I}^{2} ,
\qquad \beta=1,\ldots,l
\ear
With $(\sigma^A) = (\phi^i, \varphi^\beta)$,
according to (\ref{3.14}),
\bear{3.31}
\phi^i = \sum_{s \in S_{e}} \alpha^i_s   \ln H_s ,
\qquad i = 1, \ldots, n ,
\\ \label{3.32}
\varphi^\beta = \sum_{s \in S_{e}} \alpha^\beta_s  \ln H_s ,
\qquad \beta=1,\ldots,l ,
\ear
and the gauge (\ref{3.1}) reads
\beq{3.33}
\gamma = \sum_{s \in S_{e}}
\alpha^0_s  \ln H_s ,
\qquad
\eeq 
where
\beq{3.34}
\alpha^0_s  := \varepsilon(I) \frac{D(I)}{2-D}  \nu_{a,I}^{2} .
\eeq
With $H_{a,I}:=H_{(a,I)}$, \rf{3.29}, \rf{3.30}, and \rf{3.34}, 
the solution of \rf{3.13} - \rf{3.16} reads
\bear{xxxxx} \nn
g&=& \biggl( 
\prod_{s \in S_{e}} 
H_s^{2 \alpha^0_s} 
\biggr)  
g^{(0)}
+ \sum_{i=1}^{n}
\biggl( 
\prod_{s \in S_{e}} 
H_s^{2 \alpha^i_s} 
\biggr)  
g^{(i)} 
\\
\label{4.1}
&=&  
\left(
\prod_{(a,I)\in S_{e}} 
H_{a,I}^{ \eps(I) 2 D(I) \nu^2_{a,I} }
\right)^{1/(2-D)}  
%\times 
\left\{ 
g^{(0)}
+  \sum_{i=1}^{n} 
\left(
\prod_{(a,I) \in S_{e}, I \ni i}
H_{a,I}^{\eps(I) 2 \nu^2_{a,I} }
\right)  
g^{(i)}  
\right\} , 
\\\nn % \label{3.36}
%\qquad
&&
\mbox{\rm with}\ 
{\rm Ric}[g^{(0)}]=0,\qquad 
{\rm Ric}[g^{(i)}] =0,\qquad i=1,\ldots n ,
\ear
\bear{4.p}
\varphi^\beta  =  \sum_{s \in S_e}
\alpha^\beta_{s} \ln H_s 
= - \sum_{(a,I) \in S_{e}}
\eps(I) C^{\beta\gamma}\lambda_{\gamma a} 
\nu^2_{a,I} \ln H_{a,I} ,
&\qquad& \beta=1,\ldots,l ,
\\
\label{4.a1}
A^{a} = \sum_{I \in \Omega_{a,e}}
\frac{\nu_{a,I}}{H_{a,I}} \tau_{I} ,
&\qquad& a \in \Delta_e ,
%\\
%\label{4.a2}
%A^{a} = 0 ,
%&\qquad& a \not \in \Delta_e ,
\ear
where forms $\tau_I$ are defined in \rf{2.16a},
parameters $\nu_s \neq 0$ and ${\lambda}_a$
satisfy the orthobrane condition (\ref{3.28}),
$H_s$ are positive harmonic functions on $M_0$,
${\rm Ric}[g^{(i)}]$ denotes the Ricci-tensor of $g^{(i)}$,
and  $\prod_{\emptyset} := 1$ is always taken as convention.
The orthobrane condition
(\ref{3.28}) reads
\bear{4.2}
D(I) + \frac{(D(I))^2}{2-D}
+ C^{\alpha\beta} {\lambda}_{\alpha a}  {\lambda}_{\beta b}
&=& - \eps(I) (\nu_{a,I})^{-2} ,
\qquad  \\
\label{4.3}
D(I \cap J) + \frac{D(I) D(J)}{2-D} +
 C^{\alpha\beta} {\lambda}_{\alpha a}  {\lambda}_{\beta b} &=& 0 ,
\qquad (a,I) \neq (b,J) ,
\ear
where $I \in \Omega_{a,e}$, $J \in \Omega_{b,e}$,
$a,b \in \Delta_e$.

Note that, for positive definite $(C_{\alpha\beta})$
(or $(C^{\alpha \beta})$) and $D_0 \geq 2$, \rf{4.2} implies
\beq{3.43}
\eps(I)=-1 ,
\eeq
for all $I \in \Omega_{a,e}$, $a \in \Delta_e$.
Then, the restriction $g_{\vert M_{I}}$ of the metric \rf{4.1}
to a membrane manifold $M_{I}$ has an odd number
of negative eigenvalues,
i.e. linearly independent time-like directions.
However, if the metric $(C_{\alpha\beta})$ in the space of scalar fields
is not positive definite,
then \rf{3.43} may be violated for sufficiently negative
$C^{\alpha\beta} {\lambda}_{\alpha a}  {\lambda}_{\beta b}<0$.
In this case a non-trivial potential $A^{a}$ may also exist
on an Euclidean $p$-brane.

%%%%%%%%%%%%%%%%%%%%%%%%%%%%%%%%%%%%%%%%%%%%%%%%%%%%%%%%%%%%%%%%%%%%%%%%
\section{Examples of exact solutions}
\setcounter{equation}{0}
%%%%%%%%%%%%%%%%%%%%%%%%%%%%%%%%%%%%%%%%%%%%%%%%%%%%%%%%%%%%%%%%%%%%%%%%
Many solutions are known by now in the literature.
Among those,  by now the ones in dimensions $D=10$
and $D=11$ are studied most extensively. 
Sec. 7.1 below demonstrates just one of the most popular examples
(cf. e.g. \cite{AR,BREJS,Ts3}.) 
In Sec. 7.2 we give explicit solutions for a $10+2$-dimensional model 
(with $2$ times) related to \cite{KKP}. 
%%%%%%%%%%%%%%%%%%%%%%%%%%%%%%%%%%%%%%%%%%%%%%%%%%%%%%%%%%%%%%%%%%%%%%%%
\subsection{Membranes in $11$-dimensional supergravity}
%%%%%%%%%%%%%%%%%%%%%%%%%%%%%%%%%%%%%%%%%%%%%%%%%%%%%%%%%%%%%%%%%%%%%%%%
Now we apply our results to the special case of the
bosonic part of $11$-dimensional supergravity \cite{CJS,SS},
given by
\beq{b.16}
S_{11} =  {S}_{11,p} + c_{11} {S}_{11,t} ,
\eeq
with a perturbative sector 
%%(cf. also \cite{CJS,SS}) 
\beq{b.1}
{S}_{11,p} :=
\frac{1}{2\kappa^{2}}
\int_{M} d^{11}z \sqrt{|g|} \{ {R}[g] -  \frac{1}{4!}  (F)^2 \} ,
\eeq
a constant $c_{11}$, and 
a topological sector 
\beq{b.1t}
{S}_{11,t} :=
\frac{1}{2\kappa^{2}}
\int_{M} A \wedge F  \wedge F 
\eeq
of generalized Chern-Simons form,
where $F  =d A$ is just a single differential $4$-form. 
Recently,  in \cite{Sm}
a generalized canonical formalism  has been applied to the topological 
quantum field theory related to \rf{b.1t}.

However, for the classical solutions we need only
to consider the perturbative sector \rf{b.1}.
Since here $|\Delta|=1$, we suppress the label of different forms.
Let us consider the case where
\beq{b.2}
D(I) = 3, \quad  I \in  \Omega_{e} ,
\eeq
which is the worldsheet dimension of electrically charged $2$-branes 
(membranes).
Since here we have no extra scalar matter fields,
a special solution of orthobrane relations
\rf{4.2} and \rf{4.3}  with \rf{b.2} is given by
\beq{b.4}
D(I \cap J) = 1, \qquad I \neq J ,
\eeq
\beq{b.6}
\eps(I) = -1 ,
\eeq
\beq{b.5}
\nu^{2}_{I}= \frac{1}{2} ,
\eeq
for all $I, J\in\Omega_{e}$. 
The metric \rf{4.1} then reads
\bear{b.8}
g&=& U_e \biggr\{g^{(0)} + \sum_{i=1}^{n}
U_{i,e} g^{(i)} \biggr\}, 
\\ \label{b.9}
U_e :=
\biggr( \prod_{I \in \Omega_{e}}
H_{I} \biggr)^{\frac{1}{3}}, \qquad
\label{b.11}
&&U_{i,e} :=
 \prod_{I \in \Omega_{e}, I \ni i }H_{I}^{-1} ,
\ear
where all functions
$H_{I}$  are  harmonic on $M_0$, and
the electric $4$-form field has the solutions
\beq{b.14}
F =  \sum_{I \in \Omega_{e}} \nu_{I}
 d\left(\frac{1}{H_{I}}\right) \wedge \tau_I .
\eeq
The multidimensional manifold \rf{2.10} 
must of course be compatible
with the intersection
rules \rf{b.4} and the signature restrictions \rf{b.6}
imposed within $\Omega_{e} \subset \Omega$.
Thus, for solutions of intersection type \rf{b.4} and signature type 
\rf{b.6} there is an upper bound $D_0\leq 6$.
Conditions \rf{b.4} and \rf{b.6} admit
solutions with $2$ or $3$ time-like intersecting 
$2$-brane worldsheets, where $D_0=6$ or $D_0=4$ respectively.
(Note that, the case
of $4$ time-like intersecting $(1+2)$-dimensional manifolds
for $D_0=2$,
is excluded here, since we used the effective 
$\sigma$-model in the harmonic gauge $f=0$.) 
In this example, Proposition 1 of Sec. 4 ensures 
that there are no extra constraints for \rf{2.9}. 
%%%%%%%%%%%%%%%%%%%%%%%%%%%%%%%%%%%%%%%%%%%%%%%%%%%%%%%%%%%%%%%%%%%%%%%%
\subsection{$2$- and $3$-branes in a $10+2$-dimensional model}
%%%%%%%%%%%%%%%%%%%%%%%%%%%%%%%%%%%%%%%%%%%%%%%%%%%%%%%%%%%%%%%%%%%%%%%%
Now we illustrate a  general solution
for a bosonic model in dimension  $D= 10+2$ (with $2$ times)
that yields a consistent  truncation to the
perturbative bosonic sector of 11-dimensional
supergravity  (see \cite{KKP}).
This is a model with $2$ different field forms, a $4$-form $F^4$
and a $5$-form $F^5$, whence we set $\Delta := \{4, 5 \}$,
such that $a=n_a$ for $a\in\Delta$.

The total action of this model is
\beq{a.0}
S_{12} =  \hat{S}_{12,p} + c_{12} {S}_{12,t} . 
\eeq
with 
a perturbative sector
\beq{a.1}
\hat{S}_{12,p} =
\frac{1}{2\kappa^{2}}
\int_{M} d^{12}z \sqrt{|g|} \{ {R}[g] 
+g^{MN} \partial_{M} \varphi \partial_{N} \varphi
- \frac{1}{4!} e^{ 2 \lambda_{4} \varphi}  (F^4)^2
- \frac{1}{5!} e^{ 2 \lambda_{5} \varphi} (F^5)^2 \} ,
\eeq
a constant $c_{12}$, and a topological sector 
\beq{a.1t}
{S}_{12,t} :=
\frac{1}{2\kappa^{2}}
\int_{M} A^5 \wedge F^4  \wedge F^4 , \quad d A^5=F^5 ,  
\eeq
of generalized Chern-Simons form.
For the classical solutions we need again only
to consider the perturbative sector \rf{a.1}.
In agreement with \cite{KKP} we set
\beq{a.2}
 \lambda_{4}^2 :=  \frac{1}{10}, \quad
 \lambda_{5} := - 2 \lambda_{4}.
\eeq
Now we consider electric type solutions with 
$\Delta_e=\Delta$. Then, 
the dimensions of the 2- and 3-brane worldsheets are
\bear{a.3}
D(I) = 3&,& \qquad  I \in  \Omega_{4,e} ,   \nn\\ 
D(I) = 4&,& \qquad  I \in  \Omega_{5,e} .  
\ear
Thus, the model describes
electrically charged $2$- and $3$-branes (carrying the electrical fields 
$F^4$ resp. $F^5$).
The orthobrane relations 
\rf{4.2} and \rf{4.3} are satisfied with
\bear{a.4}
D(I \cap J) = &1, \quad \{ D(I), D(J)\} =& \{ 3, 3\}, \{ 3, 4\},  \nn\\ 
D(I \cap J) = &2, \quad \{ D(I), D(J)\} =&  \{ 4, 4\} ,
\ear
for all $I\neq J$, 
\bear{a.6}
\eps(I) = -1, 
\ear
with orthobrane coefficients
\beq{a.5}
 \nu^{2}_{4,e,I} = \nu^{2}_{5,e,I}=  \frac{1}{2} ,
\quad I \in  \Omega_{4,e} \cup \Omega_{5,e} .
\eeq
With \rf{a.6}, all electric $p$-branes must have
an odd number of time directions. 

With $H_{4,e,I}$,  
$I \in \Omega_{4,e}$,
and $H_{5,e,J}$, 
$J \in \Omega_{5,e}$,
harmonic functions on $M_0$, 
we can now describe the general solution of this model.
The metric reads
\bear{a.8}
g= U_e  \biggr\{g^{(0)} + \sum_{i=1}^{n}
U_{i,e} g^{(i)} \biggr\}, 
\\
\label{a.9}
U_e :=   \biggr( \prod_{I \in \Omega_{4,e}}
H_{4,e,I} \biggr)^{\frac{3}{10}}
\biggr( \prod_{J \in \Omega_{5,e}}
H_{5,e,J} \biggr)^{\frac{2}{5}},   
\\
\label{a.10}
U_{i,e} :=
\biggr( \prod_{I \in \Omega_{4,e}, \ I \ni i }
H_{4,e,I}^{-1} \biggr)
\biggr( \prod_{J \in \Omega_{5,e}, \ J \ni i }
H_{5,e,J}^{-1} \biggr) .
\ear
The scalar field is
\bear{a.13}
\varphi =
- \frac{\lambda_4}{2}
\sum_{I \in \Omega_{4,e}}  \ln H_{4,e,I}
+ \lambda_4 \sum_{J \in \Omega_{5,e}}  \ln H_{5,e,J} .  
\ear
The field forms are given as
\bear{a.14}
F^{4}= \sum_{I \in \Omega_{4,e}} \nu_{4,e,I}
 dH_{4,e,I}^{-1} \wedge \tau(I) , 
\\
\label{a.15}
F^{5}= \sum_{J \in \Omega_{5,e}} \nu_{5,e,J}
 dH_{5,e,J}^{-1} \wedge \tau(J) .
\ear
In the following we specify some explicit solutions
of the general shape of \rf{a.8} - \rf{a.15}
with $2$ different times.   
The solutions can be characterized by the sets
$\Omega_{a,e}, a = 4,5$. 

Let us now consider the case where $n=8$, 
the manifolds $(M_0,g^{(0)})$
and $(M_8,g^{(8)})$  having dimensions
$D_0$ resp. $D_8$ with  $D_0 + D_8 = 5$, while 
\bear{a.16}
&&M_1 = \ldots = M_7 = {\R}, \\
\label{a.17}
&&g^{(1)} = - dt_1 \otimes dt_1, \quad g^{(2)} = - dt_2 \otimes dt_2,  \\
\label{a.18}
&&g^{(i)} = dy_i \otimes dy_i, \quad i = 3, \ldots, 7.
\ear
For $D_0 = 3,4,5$, the internal space $M_8$ has corresponding
dimension $D_8 = 2,1,0$, respectively.
In order to avoid additional constraints for  \rf{2.9} 
in this example, we demand that the  orthobranes also satisfy
\rf{2.r2} of Proposition 2 in Sec. 4, and specify
some explicit examples for an  index set $S_e$
consistent with \rf{2.r2}, \rf{a.4}, and \rf{a.6}.
\nl
%%%%%%%%%%%%%%%%%%%%%%%%%%%%%%%%%%%%%%%%%%%%%%%%%%%
\noindent
{\bf Solutions with three $2$-branes and one $3$-brane:}
%%%%%%%%%%%%%%%%%%%%%%%%%%%%%%%%%%%%%%%%%%%%%%%%%%%
\nl
In a first case, we consider
\bear{a.19}
\Omega_{4,e} :=& \{A_1, A_2, A_3 \}, 
&\quad 
A_1 := \{1,3,5 \}, A_2 := \{1,4,6 \}, 
A_3 := \{2,5,6 \} , 
\\\label{a.20}
\Omega_{5,e} :=& \{ B \}, \qquad\quad 
&\quad 
B := \{2,3,4,7 \} . 
\ear
The sets \rf{a.19} and \rf{a.20} satisfy 
conditions  \rf{2.r2}, \rf{a.4} and \rf{a.6}.
Hence, a solution is given in terms of 
harmonic functions  
$H_{A_i} := H_{4,e,A_i}$, $i =1,2,3$,
and  $H_B := H_{5,e,B}$
on $(M_0, g^{(0)})$.
Its metric reads 
\bear{a.21}
g&=& 
\biggr( H_{A_1} H_{A_2} H_{A_3} \biggr)^{\frac{3}{10}}
\biggr( H_{B} \biggr)^{\frac{2}{5}}    
  \biggr\{ g^{(0)} 
-  H_{A_1}^{-1} H_{A_2}^{-1} dt_1 \otimes dt_1
-  H_{A_3}^{-1} H_{B}^{-1} dt_2 \otimes dt_2 
\nn\\
&&
\qquad\qquad\qquad\qquad\qquad
+  H_{A_1}^{-1} H_{B}^{-1} dy_3 \otimes dy_3
+  H_{A_2}^{-1} H_{B}^{-1} dy_4 \otimes dy_4 
\nn\\
&&
+  H_{A_1}^{-1}  dy_5 \otimes dy_5
+  H_{A_2}^{-1}  dy_6 \otimes dy_6 
+  H_{B}^{-1}    dy_7 \otimes dy_7  + g^{(8)} \biggr\} ,
\ear
while the scalar field is given as
\beq{a.22}
\varphi = -\frac{\lambda_4}{2}
 \ln \frac{ H_{A_1} H_{A_2} H_{A_3} }{  H_{B}^{2} }  ,
\eeq
and the field forms are 
\bear{a.23}
F^{4}&=& 
\nu_{A_1} dH_{A_1}^{-1} \wedge dt_1 \wedge dy_3 \wedge dy_5  
+ \nu_{A_2} dH_{A_2}^{-1} \wedge dt_1 \wedge dy_4 \wedge dy_6 
\nn\\
&&
+ \nu_{A_3} dH_{A_3}^{-1} \wedge dt_2 \wedge dy_5 \wedge dy_6 ,  
\nn\\
\label{a.24}
F^{5}&=& 
\nu_{B} dH_{B}^{-1} \wedge dt_2 \wedge dy_3 \wedge dy_4 \wedge dy_7 ,
\ear
with $\nu_{I}^2 = 1/2$ for all $I = A_1, A_2, A_3, B$.

The solution describes two $2$-branes with common time $t_1$
(e.g. "our" time), each intersecting another 2-brane,
which  shares a second time $t_2$
(which e.g. may be interpreted as an "internal" or "shadow" time) 
with a 3-brane, which intersects all $2$-branes.
\nl
%%%%%%%%%%%%%%%%%%%%%%%%%%%%%%%%%%%%%%%%%%%%%%%%%%%
\noindent
{\bf Solutions with two $2$-branes and two $3$-branes:}
%%%%%%%%%%%%%%%%%%%%%%%%%%%%%%%%%%%%%%%%%%%%%%%%%%%
\nl
As a second case, we consider 
\bear{a.25}
\Omega_{4,e} :=& \{ A_1, A_2 \} ,
&\quad 
A_1 := \{1,3,5 \}, \ A_2 := \{1,4,6 \},  
\\\label{a.26}
\Omega_{5,e} :=& \{ B_1, B_2 \} , 
&\quad 
B_1 := \{2,3,6,7 \}, \ B_2 := \{2,4,5,7 \} . 
\ear
The sets \rf{a.25} and \rf{a.26} satisfy  
conditions  \rf{2.r2}, \rf{a.4} and \rf{a.6}.
Hence, a solution is given in terms of 
harmonic functions  
$H_{A_i} := H_{4,e,A_i}$ and $H_B := H_{5,e,B}$, $i =1,2$,
on $(M_0, g^{(0)})$.
Its metric reads 
\bear{a.27}
g&=& 
\biggr( H_{A_1} H_{A_2}  \biggr)^{\frac{3}{10}}
\biggr( H_{B_1} H_{B_2} \biggr)^{\frac{2}{5}} 
\biggr\{ g^{(0)} 
-  H_{A_1}^{-1} H_{A_2}^{-1} dt_1 \otimes dt_1
-  H_{B_1}^{-1} H_{B_2}^{-1} dt_2 \otimes dt_2 
\nn\\
&&
\qquad\qquad\qquad\qquad\qquad
+  H_{A_1}^{-1} H_{B_1}^{-1} dy_3 \otimes dy_3
+  H_{A_2}^{-1} H_{B_2}^{-1} dy_4 \otimes dy_4 
\nn\\
&&
+  H_{A_1}^{-1} H_{B_2}^{-1} dy_5 \otimes dy_5
+  H_{A_2}^{-1} H_{B_1}^{-1} dy_6 \otimes dy_6 
+  H_{B_1}^{-1} H_{B_2}^{-1}  dy_7 \otimes dy_7  + g^{(8)} \biggr\} ,
\ear
while the scalar field is given as
\beq{a.29}
\varphi = -\frac{\lambda_4}{2}
 \ln \frac{ H_{A_1} H_{A_2} }{ H_{B_1}^{2} H_{B_2}^{2} }  , 
\eeq
and the field forms are 
\bear{a.30}
F^{4}&=& 
\nu_{A_1} dH_{A_1}^{-1} \wedge dt_1 \wedge dy_3 \wedge dy_5  
+ \nu_{A_2} dH_{A_2}^{-1} \wedge dt_1 \wedge dy_4 \wedge dy_6 ,
\nn\\
\label{a.31}
F^{5}&=& 
\nu_{B_1} dH_{B_1}^{-1} \wedge dt_2 \wedge dy_3 \wedge dy_6 \wedge dy_7
+ \nu_{B_2} dH_{B_2}^{-1} \wedge dt_2 \wedge dy_4 \wedge dy_5 \wedge dy_7 ,
\ear
where $\nu_{I}^2 = 1/2$ for all $I = A_1, A_2, B_1, B_2$.

The solution describes two $2$-branes with common time $t_1$,
each intersecting with two $3$-branes, which  share another time $t_2$
and also intersect besides.

Finally, it should also be noted that certain permutations of factor spaces,
like e.g. the permutations $(3,5)$ or $(4,6)$ in the last example, 
can yield likewise solutions.
%%%%%%%%%%%%%%%%%%%%%%%%%%%%%%%%%%%%%%%%%%%%%%%%%%%%%%%%%%%%%%%%%%%%%%%%
\section{Conclusion }
\setcounter{equation}{0}
%%%%%%%%%%%%%%%%%%%%%%%%%%%%%%%%%%%%%%%%%%%%%%%%%%%%%%%%%%%%%%%%%%%%%%%%
Above, we extended the multidimensional $\sigma$-model 
\cite{RZ,IM,Ra2} to include, besides scalar fields, generalized 
composite electric fields 
on intersecting $p$-branes compatible with its structure.
We further elaborated an ansatz for general electric $p$-brane 
solutions (see also \cite{IM4}) in terms of several harmonic 
functions on an effective external manifold $M_0$.
The criterion for solutions presented in Sec. 6 is essentially based
on mainly topological intersection conditions,
corresponding to the {\em orthobrane} property \rf{3.19}.  
  
Our orthobrane solutions generalize the usual intersecting
$p$-brane solutions to the case of Ricci-flat internal spaces
$(M_{i}, g^{(i)})$, $i=1,\ldots, n$, of arbitrary signatures. 

The solutions presented here contain 
the special case of flat factor spaces $M_{i}$, $i = 0, \ldots, n$ 
(see, for example, \cite{Ts1}).
The usually known intersecting $p$-brane solutions 
are given by (\ref{4.1}) with flat 
$M_{i} = {\R}^{D_{i}}$, where just one of them, say $M_1$ 
is pseudo-Euclidean,
$g^{(1)} = \eta_{m_{1} n_{1}} dy_1^{m_{1}} \otimes dy_1^{n_{1}}$,
and all others Euclidean,
$g^{(0)} = \delta_{\mu \nu} dx^{\mu} \otimes dx^{\nu}$,
$g^{(i)} = \delta_{m_{i} n_{i}} dy_i^{m_{i}} \otimes dy_i^{n_{i}}$ ($i > 1$),
and harmonic functions
\bear{8.1}
H_s(x) = 1 + \sum_{k = 1}^{N_s} \frac{ 2m_{s,k}}{|x - x_{s,k}|^{D_0 - 2}} ,
\qquad s \in S_e ,
\nonumber
\ear
each with poles of mass $m_{s,k}$ at $x_{s,k}$, $k=1, \ldots,N_s$,
where $1\in I$ and $D(I)=p+1$ for all $I\in \Omega_{a,e}$, $a\in\Delta_e$.
In this case the worldsheets of all $p$-branes
have a common intersection in $M_1$ containing the time submanifold.

However, in our approach there might well exist cases where (some of)
the submanifolds 
$M_I$ of (\ref{2.20}) are non-intersecting,
or contain different time submanifolds, i.e. $p$-branes
may live in different times.
In fact, recently interest has risen to consider also
solutions with arbitrary signatures \cite{AIR,AIV,IM0}.
%(see also \cite{AVD}).
In this aspect, our general solution, and the special examples of
Sec. 7.2, may be considered as a starting point for 
generalizing multi-temporal spherically-symmetric solutions
\cite{IM5} to the $p$-brane case.
We note that, in the presence  of additional scalar matter fields,  
the condition \rf{4.2} may also be consistent with 
Euclidean orthobranes.

For composite field forms of electric type \rf{2.22},
the effective Einsteinian $\sigma$-model \rf{3.2}
(with $D_0\neq 2$ in the harmonic gauge) 
admits some general solutions
in terms of harmonic functions 
for the case of vanishing dilatonic potential $V$,
i.e. Ricci flat internal spaces and vanishing cosmological constant.  
In our examples of Sec. 7, the common time of any intersecting  $p$-branes
is an internal one, not associated with the external space $M_0$,
which exclusively triggers the dynamics. Therefore, all these 
$p$-branes are static.

The general nonlinear $\sigma$-model \rf{2.1} admits also 
more general field forms than the electric ones given above by  \rf{2.22}.
In particular, it is possible to include both, electric and magnetic types.
For this more general case, static, spherically symmetric 
solutions were obtained recently in \cite{BKR}. Their properties
show some characteristic dependence 
on the intersection geometry of the $p$-branes.

The importance
of solutions with $p$-branes in the more specific  cosmological context
is also supported by recent examinations of \cite{LMP} - \cite{GrIM}. 
Even more, the Wheeler-de Witt equation and further additional constraints
(which appear only when neither \rf{2.r1} nor \rf{2.r2} holds)
are now a subject of investigation \cite{IKR}.
However, at present, $p$-brane cosmology 
still leaves many fundamental questions to be answered,
especially  concerning in relation to quantum theory.
Some future work on the $\sigma$-model also has to provide a complete 
treatment of the reparametrization gauge of the metric on $M_0$,
related to the physical question of the appropriate frame. 

However,  previous examinations at least give evidence 
that the multidimensional geometry
is a powerful ansatz for obtaining $p$-brane solutions in
the bosonic sectors of modern supergravity theories. 
\nl
\nl 
\noindent
\subsection*{Acknowledgements}
%\nl
\nl
This work was supported 
by DFG grants 436 RUS 113/7, 436 RUS 113/236/0(R), Schm 911/6 (M.R.) 
and the Russian Ministery of Science and Technology,  
Russian Fund for Basic Research, project N 95-02-05785-a (V.I., V.M.). 
V.I. and V.M. are grateful 
for hospitality at AIP and Potsdam University, 
while M.R. thanks the Russian Gravitational Society for 
the friendly care during his visit there,
when this work was completed.
The authors also thank A. Zhuk for some discussion and references. 
\np


\small
\begin{thebibliography}{99}

\bibitem{Dab}
A. Dabholkar, G. Gibbons, J.A. Harvey, and F. Ruiz Ruiz,
{\it Nucl. Phys. B} {\bf 340}, (1990) 33.

\bibitem{CHS}
C.G. Callan, J.A. Harvey and A. Strominger,
{\it Nucl. Phys. B} {\bf 359} (1991) 611;  {\it Nucl. Phys. B}
{\bf 367} (1991) 60.

\bibitem{DS}
M.J. Duff and K.S. Stelle, {\it Phys. Lett. B} {\bf 253} (1991) 113.

\bibitem{HS}
G.T. Horowitz and A. Strominger,
{\it Nucl. Phys. B} {\bf 360} (1991) 197.

\bibitem{Guv}
R. G\"{u}ven, {\it Phys. Lett. B} {\bf 276} (1992) 49.

\bibitem{KLO}
R. Kallosh, A. Linde, T. Ortin, A. Peet and A. van Proeyen,
{\it Phys. Rev. D} {\bf 46} (1992) 5278.

\bibitem{LPSS}
H. L\"{u}, C.N. Pope, E. Sezgin and K. Stelle,
{\it Nucl. Phys. B} {\bf 456} (1995) 669.

\bibitem{DKL}
M.J. Duff, R.R. Khuri and J.X. Lu,
{\it Phys. Rep.} {\bf 259}  (1995) 213.

\bibitem{GHT}
G.W. Gibbons, G.T. Horowitz and P.K. Townsend,
%{\it Preprint} hep-th/9410073.
{\it Class. Quant. Grav.} {\bf 12} (1995) 297.

\bibitem{SV}
A. Strominger and C. Vafa, 
%"Microscopic Origin of the Bekenstein-Hawking Entropy", 
%{\it Preprint}  hep-th/9601029.
{\it Phys. Lett. B} {\bf 379} (1996) 99.

\bibitem{PT}
G. Papadopoulos and P.K. Townsend, 
%"Intersecting M-branes", {\it Preprint} hep-th/9603087;
{\it Phys. Lett. B} {\bf 380}  (1996) 273-279;
{\it Phys. Lett. B} {\bf 393}  (1997) 59-64.

\bibitem{BRO}
E. Bergshoeff, M. de Roo, and T. Ortin,
{\it Phys. Lett. B} {\bf 386}  (1996) 85.

\bibitem{BKO}
E. Bergshoeff, R. Kallosh and T. Ortin,
"Stationary axion/dilaton solutions and supersymmetry",
{\it Preprint} hep-th/9605059.

\bibitem{KKLP}
N. Khvengia, Z. Khvengia, H. L\"u, C.N. Pope,
%"Intersecting M-Branes and Bound States"
%{\it Preprint} hep-th/9605077.
{\it Phys. Lett. B} {\bf 388}  (1996) 21.

\bibitem{Ts1}  
A.A. Tseytlin, 
%"Harmonic Superpositions of M-branes", {\it Preprint} hep-th/9604035; 
{\it Nucl. Phys. B} {\bf 475} (1996) 149.
%\bibitem{Ts2}
%A.A. Tseytlin,  {\it Preprint}  hep-th/9601177.

\bibitem{CM}
C.G. Callan and J.M. Maldacena, 
%"$D$-Brane Approach to Black Hole Quantum Mechanics", 
%{\it Preprint}  PUPT-1591, hep-th/9602043.
{\it Nucl. Phys. B} {\bf 472} (1996) 591-610.

\bibitem{KT}
I.R. Klebanov and A.A. Tseytlin, 
{\it Nucl. Phys. B} {\bf 475} (1996) 164;
%%%%%%%%%%%%%
%"Intersecting $M$-branes as Four-Dimensional Black Holes", 
%{\it Preprint} PUPT-1616, Imperial/TP/95-96/41,  hep-th/9604166; 
{\it Nucl. Phys. B} {\bf 475}  (1996) 179.

\bibitem{GKT}
J.P. Gauntlett,  D.A. Kastor, and J. Traschen,
%"Overlapping Branes in M-Theory", {\it Preprint} hep-th/9604179.
{\it Nucl. Phys. B} {\bf 478} (1996) 544-560.

\bibitem{LPX}
H. L\"u, C.N. Pope, and K.W. Xu, 
"Liouville and Toda solitons im M-theory",
{\it Preprint} hep-th/9604058.

\bibitem{LPS1}
H. L\"u, C.N. Pope, and K.S. Stelle,
%"Weyl Group Invariance and p-brane Multiplets",
{\it Nucl. Phys. B} {\bf 476} (1996) 89;
%\bibitem{LPS2}
%H. L\"u, C.N. Pope, and K.S. Stelle,
%"Vertical Versus Diagonal Reduction for p-Branes", 
%{\it Preprint} hep-th/9605082.
{\it Nucl. Phys. B} {\bf 481} (1996) 313-331.

%\bibitem{CH}
%M. Cvetic and C.M. Hull, {\it Preprint}  hep-th/9606193.

\bibitem{CGa}
G. Cl\'ement and D.V. Gal'tsov,
%"Stationary BPS solutions to dilaton-axion gravity"
%{\it Preprint} GCR-96/07/02 DTP-MSU/96-11, hep-th/9607043.
{\it Phys. Rev. D} {\bf 54} (1996) 6136-6152.

\bibitem{IM4}
V.D.Ivashchuk and V.N.Melnikov,
%"Intersecting p-brane Solutions in Multidimensional Gravity and M-theory", 
%{\it Preprint}  hep-th/9612089;
{\it Gravitation and Cosmology} {\bf 2} (1996) 297-305.

\bibitem{V}
A. Volovich, "Three-block $p$-branes in various dimensions",
{\it Preprint} hep-th/9608095.

\bibitem{AV}
I. Aref'eva and A. Volovich, "Composite $p$-branes in diverse dimensions",
{\it Preprint} SMI-19-96, hep-th/9611026.

\bibitem{AR}
I.Ya. Aref'eva and O.A. Rytchkov,
"Incidence matrix description of intersecting $p$-brane solutions",
{\it Preprint} SMI-25-96, hep-th/9612236.

\bibitem{AVV}
I.Ya. Aref'eva, K. Viswanathan and I.V. Volovich,
%"p-Brane Solutions in Diverse Dimensions",
%{\it Preprint} hep-th/9609225.
{\it Phys. Rev. D} {\bf 55} (1997) 4748-4755.

\bibitem{AVV}
I.Ya. Aref'eva, K. Viswanathan, A.I. Volovich and I.V. Volovich,
%"p-brane solutions in diverse dimensions",
%{\it Preprint} hep-th/9701092.
{\it Nucl. Phys. Proc. Suppl.} {\bf 56B} (1997) 52-60.

\bibitem{BBJ}
K. Behrendt, E. Bergshoeff, and B. Janssen,
{\it Phys. Rev. D} {\bf 55} (1997) 3785-3796.

\bibitem{BRP}
E. Bergshoeff, M. de Roo, and S. Panda,
{\it Phys. Lett. B} {\bf 390}  (1997) 143-147.

\bibitem{AEC}
R. Argurio, F. Englert and L. Hourant, 
"Intersection rules for $p$-branes", hep-th/9701042.

\bibitem{AIR}
I.Ya. Aref'eva, M.G. Ivanov and O.A. Rytchkov,
"Properties of intersecting $p$-branes in various dimensions",
{\it Preprint} SMI-05-97, hep-th/9702077.

\bibitem{LP}
H. L\"u and C.N. Pope, 
"$p$-brane taxonomy", hep-th/9702086.

\bibitem{S}
K.S. Stelle, 
"Lectures on supergravity $p$-branes", hep-th/9701088.

\bibitem{M}
S.D. Majumdar, 
{\it Phys. Rev.} {\bf 72} (1947) 930.

\bibitem{P}
A. Papapetrou,  
{\it Proc. R. Irish Acad. A} {\bf 51} (1947) 191.

\bibitem{CJS}
E. Cremmer, B. Julia, and J. Scherk, 
{\it Phys. Lett. B} {\bf 76} (1978) 409.

\bibitem{SS}
A. Salam and E. Sezgin, 
eds., "Supergravities in diverse dimensions", 
Reprints in 2 vols., World Scientific (1989).
%
%\bibitem{GSW}
%M.B. Green, J.H. Schwarz, and E. Witten, "Superstring
%Theory" in 2 vols. (Cambridge Univ. Press, 1987).

\bibitem{HT}
C. Hull and P. Townsend, {\it Nucl. Phys. B} {\bf 438} (1995) 109.

\bibitem{T}
P.K. Townsend, {\it Phys. Lett. B} {\bf 350} (1995) 184.

\bibitem{Wit}
E. Witten,  
{\it Nucl. Phys. B} {\bf 443} (1995) 85.

\bibitem{HW}
P. Horava and E. Witten, {\it Nucl. Phys. B} {\bf 460} (1996) 506. 
%{\it Preprint} hep-th/9603142; 

\bibitem{D}
M.J. Duff,  
%"M-theory (the Theory Formerly Known as Strings)",
%{\it Preprint} CTP-TAMU-33/96, hep-th/9608117.
{\it Int. J. Mod. Phys. A} {\bf 11} (1996) 5623-5642.

%M.J. Duff, J.T. Liu, and R. Minasian, 
%"Eleven-dimensional Origin of String-String Duality: a One-Loop Test", 
%{\it Preprint} hep-th/9506126.

\bibitem{Sch}
J.M. Schwarz,  "Lectures on Superstring and M-theory Dualities",
{\it Preprint} ICTP, hep-th/9607201.

\bibitem{IMZ}
V.D. Ivashchuk, V.N. Melnikov and A. I. Zhuk,
{\it Nuovo Cimento B} {\bf 104} 575 (1989).

\bibitem{IM3}
V.D. Ivashchuk and V.N. Melnikov,
{\it Int. J. Mod. Phys. D} {\bf 3} (1994) 795; 
{\it Gravitation and Cosmology} {\bf 1} (1995) 204. 

\bibitem{Mel}
V.N. Melnikov,
"Multidimensional Cosmology and  Gravitation",  
{\it Preprint} CBPF, Rio de Janeiro (1995).

\bibitem{Ra1}
M. Rainer,
{\it Int. J. Mod. Phys. D} {\bf 4} (1995) 397-416,
{\it Gravitation and Cosmology} {\bf 1} (1995) 121-130.

\bibitem{GIM}
V.R. Gavrilov, V.D. Ivashchuk, and V.N. Melnikov,
{\it J. Math. Phys} {\bf 36}, (1995) 5829. 

\bibitem{GKMR}
V.R. Gavrilov, U. Kasper, V.N. Melnikov, and M. Rainer,
"Toda chains with type $A_m$ Lie algebra for 
multidimensional $m$-component perfect fluid cosmology",
{\it Preprint} Univ. Potsdam (1997).

\bibitem{RZ}
M. Rainer and A. Zhuk, {\it Phys. Rev. D}  {\bf 54} (1996) 6186.

\bibitem{IM}
V.D. Ivashchuk and V.N. Melnikov,
%"Multidimensional Gravity with Einstein Internal spaces",
%preprint RGS-96-003, hep-th/9612059; 
{\it Gravitation and Cosmology} {\bf 2} (1996) 177.

\bibitem{Ra2}
M. Rainer, 
"Effective multi-scalar-tensor theories and 
$\sigma$-models from multidimensional gravity",
{\it Preprint} P-Math-97/4, Univ. Potsdam (1997);
"Multidimensional scalar-tensor theories and 
minisuperspace approach", to appear in: 
{\it Proc. Int. Workshop on 
Modern Modified Theories of Gravitation and Cosmology},
Beer Sheva (1997).

\bibitem{GH}
G.W. Gibbons and S.W. Hawking, 
{\it Phys. Rev. D} {\bf 15}, 2752 (1977).

\bibitem{Y}
J.W. York, 
{\it Phys. Rev. Lett.} {\bf 28}, 1082-1085 (1972);
{\it Found. Phys.} {\bf 16}, 249-257 (1986).

\bibitem{BREJS}
E. Bergshoeff, M. de Roo, E. Eyras, B. Janssen, and J.P. van der Schaar, 
%"Multiple intersections of $D$-branes and $M$-branes",
%{\it Preprint} hep-th/9612095.
{\it Nucl. Phys. B} {\bf 494} (1997) 119-143.

\bibitem{Ts3}
A.A. Tseytlin, 
%"Composite BPS configurations of p-branes in 10 and 11 dimensions",
%{\it Preprint} hep-th/9702163.
{\it Class. Quant. Grav.} {\bf 14} (1997) 2085-2105.

\bibitem{KKP}
N. Khvengia, Z. Khvengia, H. L\"u, and  C.N. Pope,
"Toward field theory of F-theory"
{\it Preprint} hep-th/9703012.

\bibitem{Sm}
L. Smolin,
Chern-Simons theory in $11$ dimensions as a non-perturbative phase 
of $M$-theory,
hep-th/9703174.

%\bibitem{AVo}
%I.Ya. Aref'eva and I.V. Volovich,
%{\it Phys. Lett. B} {\bf 164}  (1985) 535;
%
%\bibitem{ADVo}
%I.Ya. Aref'eva, B. Dragovich and I.V. Volovich,
%{\it Phys. Lett. B} {\bf 177} (1986) 357.

\bibitem{AIV}
I.Ya. Aref'eva, M.G. Ivanov, and I.V. Volovich,
%"Non-extremal Intersecting p-branes in Various Dimensions",
%{\it Preprint} SMI-06-97, hep-th/9702079.
{\it Phys. Lett. B} {\bf 406} (1997) 44-48.

\bibitem{IM0}
V.D. Ivashchuk and V.N. Melnikov,
"Multidimensional classical and quantum cosmology 
with intersecting $p$-branes", 
{\it Preprint}  hep-th/9708157.

\bibitem{IM5}
V.D. Ivashchuk  and  V.N. Melnikov,
{\it Class. and Quant. Grav.} {\bf 11} (1994) 1793.

\bibitem{BKR}
K.A. Bronnikov, U. Kasper, and M. Rainer,
"Intersecting electric and magnetic $p$-branes:
spherically symmetric solutions",
{\it Preprint} P-Math-97/21, gr-qc/9708058.

\bibitem{LMP}
H. L\"u, S. Mukherji, and C.N. Pope, 
"From $p$-branes to cosmology",
{\it Preprint} hep-th/9612224.

\bibitem{LMMP}
H. L\"u, J. Maharana, S. Mukherji, and C.N. Pope, 
"Cosmological solutions, $p$-branes, and the Wheeler-de Witt equation",
{\it Preprint} hep-th/9707182.

\bibitem{BGrIM}
K.A. Bronnikov, M.A. Grebeniuk, V.D. Ivashchuk,  and  V.N. Melnikov,
%"Integrable multidimensional cosmology 
%for intersecting p-branes".
{\it Gravitation \& Cosmology} {\bf 3} (1997) 1.

\bibitem{GrIM}
M.A. Grebeniuk, V.D. Ivashchuk,  and  V.N. Melnikov,
"Integrable multidimensional classical and quantum cosmology 
for intersecting $p$-branes", 
{\it Preprint}  gr-qc/9708031.

\bibitem{IKR}
V.D. Ivashchuk, U. Kasper, and M. Rainer,
"Constraints in a $\sigma$-model with intersecting composite $p$-branes",
{\it Preprint} Univ. Potsdam (1997).
\end{thebibliography}
\end{document}